\documentclass{article}

\usepackage{indentfirst}
\usepackage{algorithm} % For algorithms
\usepackage{amsmath} % For mathematical equations
\usepackage{graphicx} % For including figures
\usepackage{cite} % For citations
\usepackage{lipsum} % For generating dummy text
\usepackage{array, makecell} % For table formatting
\usepackage{subcaption} % For subfigures
\usepackage{geometry}

\newgeometry{vmargin={15mm}, hmargin={17mm,17mm}}   % set the margins
\title{Propius: A Platform for Collaborative Machine 
Learning across the Edge and the Cloud}
\author{Eric Ding, Cornell University}
\date{\today}

\begin{document}

\maketitle

\def\name{{Propius}}
\def\topic{{collaborative ML}}

\begin{abstract}
    % Abstract of your paper
    % TODO add supports
    Collaborative Machine Learning is a paradigm in the field of distributed machine learning, designed to address the challenges of data privacy, communication overhead, and model heterogeneity.
    There have been significant advancements in optimization and communication algorithm design and ML hardware that enables fair, efficient and secure {\topic} training.
    However, less emphasis is put on {\topic} infrastructure development. Developers and researchers often build server-client systems for a specific {\topic} use case, which is not scalable and reusable.
    As the scale of {\topic} grows, the need for a scalable, efficient, and ideally multi-tenant resource management system becomes more pressing.
    We propose a novel system, {\name}, that can adapt to the heterogeneity of client machines, and efficiently manage and control the computation flow between ML jobs and edge resources in a scalable fashion.
    {\name} is comprised of a control plane and a data plane. The control plane enables efficient resource sharing among multiple {\topic} jobs and supports various resource sharing policies, while the data plane improves the scalability of {\topic} model sharing and result collection. Evaluations show that {\name} outperforms existing resource management techniques and frameworks in terms of resource utilization (up to $1.88\times$), throughput (up to $2.76\times$), and job completion time (up to $1.26\times$).
    %TODO data
\end{abstract} 

\section{Introduction}
\label{sec:introduction}

Collaborative Machine Learning is a paradigm in the field of distributed machine learning, 
designed to address the challenges of data privacy issues.
Computation tasks are distributed over client edge devices 
(mobile phones, IoT devices) and silos 
(companies, hospitals) for machine learning and data analysis. Collaborative ML
can be viewed as a superset of Federated Learning (FL) \cite{mcmahan2017communication}.
%TODO explain relationship between collaborative ML and FL
%TODO more use case/background for collaborative ML

Collaborative ML enables privacy-preserving and affordable ML, and unlocks huge amounts of private data for training by moving 
the models to the client side. These benefits are becoming more prominant in the age of generative AI, 
as limited public-accessible data are available for 
generative model training, 
and concerns about privacy are growing. With {\topic}, developers can have the access to an increasing amount of private data for building better models in terms of accuracy and customization.
As main {\topic} workload is performed on client devices with client data, it is also natural for clients to continue use trained local model for inference with better customization and privacy.

However, the large volume and heterogeneity nature of the 
client machines lead to challenges in device resource management. Collaborative ML developers usually need to build custom server-client protocols and softwares to implement custom algorithm to enforce client selection logic, which is time-consuming and cost-prohibitive.
If individual job framework requires a fixed set of resource due to the lack of resource management, client resources could be underutilized, as some jobs may not be able to fully utilize the resources allocated to them.
Resource sharing could be beneficial in improving resource utilization, but it could also lead to resource contention and thus long training time, as multiple {\topic} jobs are running on the same client corpus, requesting resources with the best performance and source data quality.

On top of the resource management issues present in {\topic}, ML plan (weights, training parameters) delivery and result collection (gradients, output of model) across large number of jobs and clients present a scalability challenge. Typical number of simultaneously participating clients required for a single job could be in the order of tens to tens of thousands. Also, low bandwidth network between client edge devices and parameter 
servers leads to long communication latency, creating model update and compute ramp-up delay.
%TODO think about llm

%TODO add existing solutions
Existing resource management techniques and frameworks 
fail to rise up to the challenge. 
Conventional resources in the cloud (GPUs in data centers, data lakes) often have better characteristics than client machines in terms of availability and capacity. Federated learning frameworks (Flame \cite{daga2024flamesimplifyingtopologyextension}, FedScale \cite{fedscale-icml22}, etc.) mostly focus on optimizing single-job training performance, and do not consider the resource management and sharing among multiple jobs.
% These frameworks also do not take into 
% account the second-sale communication latency between parameter 
% servers and client machines.

To solve these problems, we propose a novel system, {\name}, that can adapt to the heterogeneity of client machines, and efficiently manage and control the computation flow between ML jobs and edge resources in a scalable fashion.

% Introducing the control plane
{\name} is comprised of a control plane and a data plane. We propose a novel control plane design where the scheduling mechanism is tailored to heterogeneous and transcient resources in {\topic} setting. 
Contrary to traditional resource scheduling system where resources are generally static and long-lived, {\name} resources, such as edge devices, are short-lived and dynamic. 
This challenge is addressed by the soft-state design principle of the scheduler, where only the states of clients in a
sliding window are maintained in the system. Most of the scheduling states and decisions are mounted on the job framework ends, whose states and scale are more stable and manageable. 
Specifically, there are two modes of scheduling in {\name} scheduler, 
which scheduling policy plugins can choose from: (1) online scheduling, and (2) small-batch scheduling.

%TODO introducing the data plane
{\name} data plane enables scalable {\topic} plan sharing and result collection for large number of jobs and clients. Optimization algorithms for {\topic}, such as FedAvg \cite{mcmahan2017communication}, are supported in {\name} data plane, as long as they satisfy associativity and communitivity.

In summary, {\name} provides several benefits: it
\begin{enumerate}
    \item Hides the details of heterogeneous client resource management so developers can focus on developing learning and analytics models and algorithms instead.
    \item Avoids allocation conflict and maximizes resource utilization by coordinating among multiple collaborative ML jobs.
    \item Allows jobs to cherry-pick desired amount of resources in terms of device system attributes.
    \item Supports various scheduling algorithms for different performance objectives.
\end{enumerate}

% {\name} is not the first resource management and orchestration platform operating at such scale, but it's one of the few that could solve unique problems present in {\topic} settings, with high degree of flexibility, resiliency and completeness.

% \section{Background and Motivation}
% \label{sec:motivation}

% \subsection{Collaborative Machine Learning}
% 1. Use case

% \subsection{Current Systems and Frameworks Design}
% % Google
% % Apple
% % Frameworks

% \subsection{Inefficiencies in Existing Systems}
% 1. \topic
% 2. CDN
% %TODO
% Edge Side Include for dynamic content generation and personalization, not suitable for ML workloads

% \begin{enumerate}
%     \item Maintain full client states in schedulers (Non FL)
%     \item Inflexiblity in scheduling policy
%     \item Not suited for geo-distributed workloads
%     \item Multi-handshake binding process (Google)
% \end{enumerate}

% \subsection{Opportunities for Improvements}
\section{{\name} Overview}
\label{sec:overview}

\begin{figure}[H]
    \centering
    \includegraphics[width=0.8\textwidth]{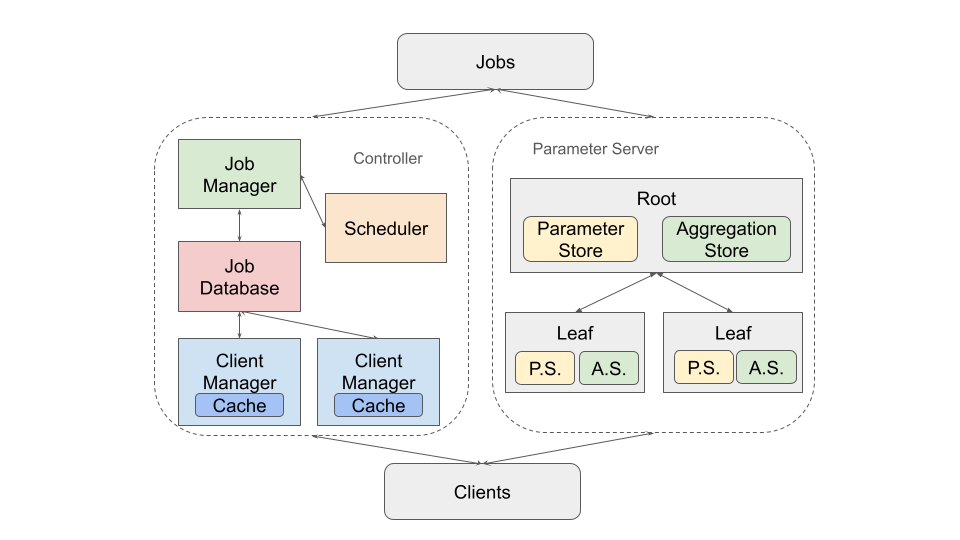} \caption{{\name} Architecture} 
    \label{fig:overview}
\end{figure}

{\name} consists of two main components: 
the control plane and the data plane.
% Control plane
The control plane is responsible for managing the job and client requests, 
and providing coordination during the resource allocation process.
% Data plane
The data plane is responsible for distributing execution planes 
from the cloud to geo-distributed clients, 
collecting updates or outputs from clients, 
and performing reduction operation during results collection. Fig. \ref{fig:overview} shows the architecture of {\name}.

% Targeted job framework and job library
{\name} is designed to support a wide range of job frameworks, as long as they fall into the category of server-client topologies and adopt round-based communication pattern where there are a one-to-all model plan distribution phase and a all-to-one reduction (aggregation) phase in each round. {\name} does not support hybrid topologies where clients communicate with each other \cite{daga2024flamesimplifyingtopologyextension}.

% Client and client library

% Workflow description and workflow diagram for job and client

% Flexibility of data plane, on or off
Through {\name} data plane, individual job framework has the access to current 
round results computed by the data plane through a simple API call. 
Each job framework needs to provide a reducer plugin for the data plane to use during the reduction phase. \
All the reduction computation from the contributions of all 
selected clients for the current round and communications are offloaded to the data plane, 
and job frameworks are only responsible for the model update step at the end of each round for better customization
and generalizability of the data plane.
However, the convenience comes at a cost. Job frameworks do not have access to individual 
client constributions for the current round as well as for the past rounds. If the job frameworks
require those information, they should turn off the data plane offloading, and use their
custom implementation to collect and perform aggregation.

%TODO scheduling problem, incorporating the graphicx
\section{{\name} Control Plane}
\label{sec:control}

{\name} controller consists of three layers: 
\ref{sec:application} application layer, 
\ref{sec:scheduling} scheduling layer, and 
\ref{sec:binding} binding layer. 
{\name} application layer handles job requests, performs admission control, 
and manages job state in the application layer. 
{\name} scheduling layer executes a scheduling policy based on real-time 
information of job states and client resources in the scheduling layer. 
It groups active jobs, and assigns priority between or within job groups. 
{\name} client manager handles client requests, 
and allocates available and eligible client resources to outstanding jobs in the binding layer. Multiple client managers can be instantiated to distribute the load from client requests.

\subsection{Application Layer}
\label{sec:application}
Job manager and job database are located in the application layer, serving 
job frameworks.
\begin{enumerate}
    \item \emph{Job framework}, or simply a job, is a specialized workload for {\topic} 
    that regulates the ML learning process and manages the 
    communication between clients. An example of a job framework is FedScale 
    parameter server \cite{fedscale-icml22}. Currently, {\name} supports job 
    frameworks that are round-based, with specific demands and constraints on 
    client resources in round granularity. Job constraints are the minimum 
    requirements on client CPU, memory, etc., and demand is the desired number 
    of clients per round. Constraints are specified by individual job frameworks, 
    and can be classified into two types: public constraints and private constraints. 
    Public constraints are enforced by {\name} upon client attributes that clients are willing to share (CPU, OS version, etc.), 
    while private constraints are enforced upon attributes that clients are reluctant to share, such as private metadata on datasets. {\name} will pass the private constraints to clients 
    for local binding decision.
    %TODO add a workflow diagram here

    \item \emph{Job manager} interfaces with jobs over WAN, 
    maintains job states, collects job metadata (number of rounds, 
    IP and port, etc.), constraints and demands on client resources in 
    the job database. Additionally, the job manager performs admission control, rejecting job requests if their demand exceeds allocation limitation during initial registration. It forwards successful job requests to the scheduling
    layer via RPC calls, which initiates scheduling, allocation and binding process. The endpoints the job manager exposes are listed in Table \ref{tab:job_manager}.

    \item \emph{Job database} keeps track of job states, and maintains job
    metadata. It is a key-value datastore indexed by job ID, accessbile to the job manager, the scheduler, and the client manager in the control plane. It also serves as the means of communication between the three control plane layers, 
    where scheduling decisions are stored by the scheduler, and retrieved by the client manager. Key fields in the job database schema are shown in 
    Table \ref{tab:job_database}.
\end{enumerate}

\begin{table}[t!]
    \centering
    \caption{Endpoints Exposed by Job Manager.}
    \begin{tabular}{|c|c|c|c|}
        \hline
        Endpoint & Description & Key Parameters & Returns\\
        \hline
        \hline
        JOB\_REGIST & Register a new job & \makecell{total\_round \\ est\_demand, \\ public\_constraint, \\ 
        private\_constraint} & \makecell{ack, \\job\_id}\\
        \hline
        JOB\_REQUEST & Request a new round & job\_id, demand & ack \\
        \hline
        JOB\_END\_REQUEST & End a round request & job\_id & ack \\
        \hline
        JOB\_FINISH & Finish a job & job\_id &  ack\\
        \hline
    \end{tabular}
    \label{tab:job_manager}
\end{table}

\begin{table}[t!]
    \centering
    \caption{Key Fields in Job Database Schema.}
    \begin{tabular}{|c|c|c|}
        \hline
        Field Name & Type & Description \\
        \hline
        \hline
        time\_stamp            & numeric & job registeration time                                                 \\
        \hline
        total\_sched           & numeric & \makecell{total time spent in waiting \\for clients over all rounds}   \\
        \hline
        start\_sched           & numeric & \makecell{last timestamp when the \\ job starts a new round}           \\
        \hline
        job\_ip                & text    & IP address of the job                                             \\
        \hline
        port                   & numeric & port number                                                       \\
        \hline
        total\_demand          & numeric & total number of clients estimated                                 \\
        \hline
        total\_round           & numeric & total number of rounds estimated                                  \\
        \hline
        attained\_service      & numeric & total number of clients attained                                  \\
        \hline
        round                  & numeric & round number                                                      \\
        \hline
        demand                 & numeric & number of clients requested in this round                            \\
        \hline
        amount                 & numeric & number of clients attained in this round                          \\
        \hline
        score                  & numeric & priority score                                                    \\
        \hline
        public\_constraint.[x] & numeric & \makecell{lower bound for client public \\ attribute value for constraint x}\\
        \hline
        private\_constraint.[y]& numeric & \makecell{lower bound for client private \\ attribute value for constraint y}\\
        \hline
    \end{tabular}
    \label{tab:job_database}
\end{table}

% {\name} returns real-time scheduling and allocation telemetry to job frameworks 
% (queueing time, allocation time, the amount of eligible resources, etc.), 
% allowing jobs to dynamically adapt to dynamic resource availability 
% and congestion level.

\subsection{Scheduling Layer}
\label{sec:scheduling}

On a high level, {\name} employs a two-level scheduling process: global scheduling, 
and client binding. {\name} scheduler is responsible for making centralized global
scheduling decisions on coarse client resource allocation. Clients can further perform
fine-grained binding decisions based on local, private attributes and the constraints
of tasks offered by {\name}.

% two modes of scheduling
{\name} should be highly scalable, capable of handling large number of clients and jobs, 
potentially in the order of millions of concurrent client tasks. It will be 
expensive to keep track of every client status in the system. On top of the scalability 
challenge, the transient and dynamic nature of clients in {\topic} settings makes it difficult to 
maintain correct client states. To ensure state correctness, the system 
needs to communicate with clients frequently, which will incur huge communication traffic. 
These problems motivate us to move away from traditional stateful design, such as Slurm \cite{yoo2003slurm}, 
to a soft-state design.

%TODO add typical job demand size

{\name} scheduler can be configured with two modes of scheduling: (1) online scheduling, and (2) 
small-batch scheduling. %TODO add plots and add descriptions for the plot
In online scheduling, {\name} does not track client states at all. Instead, scheduling policy plugins pre-compute
a priority score for every job, and store the scores in the shared job database. This action can be triggered
by job registeration or job request event, or asynchronously, which is defined by scheduling policies. Policy plugins 
can compute the score from job metadata, and global request and demand statistics by calling the set of APIs exposed by
{\name} scheduler service (Table \ref{tab:scheduling_api}). After the scores are stored in the shared job database, 
client manager will pair newly-arrived clients to requesting jobs with the highest scores while ensuring the job public constraints 
are satisfied, and send job private constraints to the clients for client-side binding.
%TODO workflow

%TODO
\begin{table}[t!]
    \centering
    \caption{Key API for Scheduling.}
    \begin{tabular}{|c|c|}
        \hline
        API & Description\\
        \hline
        \hline
        job\_db\_portal.exist(job\_id) & 
        Query whether a job exists in the system \\
        \hline
        job\_db\_portal.get\_job\_size() & 
        Get the number of jobs in the system \\
        \hline
        job\_db\_portal.get\_field(job\_id, ``[field\_name]'') & 
        Get the field value of a job \\
        \hline
        job\_db\_portal.query(query\_str) & 
        Get a list of jobs that satisfy query string \\
        \hline
        job\_db\_portal.set\_score(score, job\_id) & 
        Set priority score for a job \\
        \hline
        client\_db\_portal.get\_client\_size() & 
        \makecell{Get the number of clients \\within the system window}\\
        \hline
        \makecell{client\_db\_portal.\\get\_client\_proportion(public\_constraint)} & 
        \makecell{Get the proportion of clients \\within the system window \\
                of which attributes satisfy the public constraint}\\
                \hline
        client\_db\_portal.get\_client\_subset\_size(query\_str) & 
        \makecell{Get the number of clients \\within the system window \\
        of which satisfy the query string}\\
        \hline
    \end{tabular}
    \label{tab:scheduling_api}
\end{table}

Online scheduling mode binds available clients to outstanding jobs as soon as 
their requests arrive. It requires a priority order amoung requesting jobs enforced 
by scheduling policy plugins so that the binding layer can make fast decisions for 
hot requests from clients. This, however, could lead to head-of-line (HOL) issues, and 
inflexible policy design due to the ordering requirement.
For example, it will be difficult to implement fair-sharing scheduling in online scheduling, 
as it requires the plugins to re-compute the entire score-based priority ordering based 
on real-time global allocation status at a high frequency. 

In small-batch scheduling mode, {\name} binding layer stores temporary states of available clients in a data cache,
from which requesting jobs can select eligible clients through querying. Requesting jobs can be grouped into 
multiple job groups by scheduling policy plugins. Each job group can specify a set of constraints on client resources
through query strings. Priority order could also be maintained within each job group by plugins.
The cache can be viewed as a sliding window of client states, where only the most recent available client states are stored.
Once the client is selected by a job, it is removed from the cache, and the cache is updated with new client states. This 
prevents requesting jobs being paired with stale clients. Through specifying querying strings for each job group, 
the policy plugin can partition the resources among job groups, avoiding HOL blocking as well as enforcing job constraints on resources at the same time.
% On a high level, one can think of the online scheduling as clients choose jobs, whereas the small-batch scheduling as jobs select clients.
This follows the partition scheduling paradigm introduced in POP \cite{narayanan2021solving},
as the scheduling problem we are solving is a granular allocation problem, where each job requests a small fraction of resources, and resources are
fungible.

The job group partitioning has other benefits. It enables more general scheduling policies by removing the necessity of pre-computing priority 
scores, and reduces overall reduced scheduling runtime. The flexiblity derives from the fact that a priority order between any two jobs is not required,
and the policy plugin can arrange these jobs in different job groups, selecting resources from different client partitions. By breaking the scheduling
problem into sub-problems, a super-linear runtime speedup can be achieved, and these sub-problems can 
be executed in parallel \cite{narayanan2021solving}. The downside of small-batch scheduling is pro-longed waiting time for avaialble clients, 
leading to higher probability of task failure as clients may become offline during the waiting. %TODO algorithm example

%other aspects like preemption, and no multi-tenency on clients, %Replicated cell, distributed scheduling
Note that both the scheduling order and the job constraints are considered in these two modes. {\name} scheduler service will enforce
the job constraints automatically without the need for explicit binding rules for policy plugins to specify. Policy plugins are 
only responsible for computing the priority order of jobs, or partitioning the resources among job groups. Preemption is not 
supported in the current design, as the resources are fungible, and the communication overhead (between the cloud and edge) is high.
Only one task will be placed onto one client resource at a time, as the resource is typically weak in computation power.
{\name} employs a two-level scheduling process. Contrary to Mesos \cite{hindman2011mesos} two-level scheduling, 
where the scheduler assigns resources to jobs frameworks and asks jobs to reject the resources if they do not meet the constraints, 
{\name} let individual client resources to select suitable jobs based on private attributes and job constraints, empowering clients 
in the binding process for {\topic} tasks and reducing round-trip traffics between individual job frameworks and clients. 

\subsection{Binding Layer}
\label{sec:binding}
In binding layer, {\name} client manager conducts the first level scheduling, binding tasks from job frameworks to client resources, 
based on the priority score (online scheduling) or group partitioning (small-batch scheduling) computed by the scheduling 
layer. Each active and free clients, recieves metadata for one or more framework tasks if eligible, and decides which task to accept
based on the task constraints on private attributes. {\name} offers a client library that can be integrated into client-side
ML application, and handles the binding process and communication with the {\name} binding layer. Client-side application can specify 
the decision logic in a plugin function that is called by the {\name} client library. After determining which task to accept, 
the client sends the information back to the binding layer. The binding layer updates the job database with the binding result, 
incrementing the allocation counter for the job, until the allocation count reaches the requested amount of resources by the job 
framework at the begining of the round.

The client manager stores client metadata in shared client database (accessible by job manager and scheduler) using a sliding window scheme, 
such as public attributes (CPU, OS version, etc.) and availability timestamps. 
Global statistics on client resources can be captured by the scheduling layer to compute the priority score for each job, 
and can be queried by individual job frameworks for dynamic task planning. Schedulers or job frameworks can call 
$client\_db\_portal.get\_client\_proportion(attributes\_threshold)$ to get an idea of the global client resource distribution and 
eligible client subset size. The metadata store is also useful for general monitoring and logging purposes. 
In addition to the client database, the client manager caches the most recently available client information in small-batch scheduling
mode. As mentioned in $\S$\ref{sec:scheduling}, scheduler plugins can provide a query for each job group for client resource partitioning.
The client manager performs the query on the client cache, and assigns job partitions to queried clients.  

To handle the large number of client requests, {\name} binding layer can be horizontally scaled to distribute the load among multiple client 
managers. A load balancer is placed in front of the binding layer. The database and cache is sharded along with every instance of the client manager.

\section{Propius Data Plane} \label{sec
}

To enhance the performance and efficiency of content delivery for \topic, we propose a content delivery policy that can be integrated into existing content delivery networks (CDNs). Our approach aims to optimize the distribution and retrieval of content, particularly in scenarios requiring dynamic updates and low-latency interactions, such as federated learning models or large-scale data processing tasks.

The core of our policy revolves around leveraging the inherent capabilities of CDNs while introducing specific mechanisms tailored to the needs of \topic. By doing so, we ensure that content delivery not only meets the demands of high availability and reliability but also aligns with the specialized requirements of \topic.

\subsection{Data Plane Protocol}

The data plane protocol for {\topic} involves several key steps to ensure efficient content delivery and processing. The protocol is designed to facilitate the distribution of model execution plans and aggregation of results while minimizing latency and optimizing resource usage. The steps are as follows:

\begin{enumerate} \item \textbf{Client Request:} A client initiates a request for an execution plan from the nearest leaf server. \item \textbf{Model Check:} The leaf server verifies if it has the requested model or execution plan in its cache. \item \textbf{Pull Caching:} If the model is not available in the leaf server, it sends a request to its parent server to retrieve the model. \item \textbf{Plan Distribution:} Upon obtaining the model, the leaf server sends the execution plan to the client and caches the plan for a specified time-to-live (TTL). \item \textbf{Result Upload:} The client processes the execution plan and uploads the results to the leaf server. \item \textbf{Result Aggregation:} The leaf server aggregates the results received from clients. \item \textbf{Periodic Update:} The leaf server periodically sends partially-aggregated results to its parent server to maintain updated information across the network. \end{enumerate}

The caching strategy within this protocol involves decisions on the content to cache, the duration of caching (TTL), and eviction policies. This ensures that frequently accessed models and execution plans are readily available while managing storage resources efficiently.

Support for aggregation algorithms such as Federated Averaging (FedAvg) \cite{mcmahan2017communication} is integral to this protocol. The protocol supports any aggregation algorithm that satisfies associativity and commutativity, allowing for flexibility in optimization methods. Optimization algorithms will be implemented by individual job frameworks, ensuring compatibility with various task requirements.

\subsection{Integration with Existing CDNs}

The proposed data plane protocol is designed to integrate with existing CDN infrastructures. It leverages the established server selection algorithms and load balancing mechanisms inherent in CDNs to achieve cost savings and enhance performance. Specifically:

\begin{enumerate} \item \textbf{Server Selection:} The protocol utilizes existing server selection algorithms to route client requests to the most appropriate leaf server. This ensures efficient handling of requests and minimizes latency. \item \textbf{Load Balancing:} By relying on the CDN’s built-in load balancing capabilities, the protocol ensures that the load is distributed evenly across the network, optimizing resource utilization and avoiding overloading individual servers. \item \textbf{Cost Efficiency:} The integration with CDNs enables cost-effective content delivery by taking advantage of the CDN's infrastructure and pricing models. This approach reduces the need for additional infrastructure investments while maintaining high performance. \end{enumerate}

In summary, our data plane protocol enhances the content delivery process for {\topic} by leveraging existing CDN infrastructure and introducing specific mechanisms to address the unique requirements of \topic. This integration ensures improved performance, cost efficiency, and adaptability to various content delivery needs.

% \section{Propius Data Plane}
% \label{sec:data}

% %TODO CDN structure
% \subsection{Efficient Content Delivery for \topic}
% We propose a content delivery policy for \topic that can be implemented
% in existing content delivery networks.

% %TODO protocol
% \subsection{Data Plane Protocol}
% 1. Client request for an execution plan
% 2. Leaf server check if it has the model
% 3. If not, request from parent (pull caching)
% 4. Leaf server send model plans to clients, cache the plan (ttl)
% 5. Client upload execution result to leaf server
% 6. Leaf server aggregate results
% 7. Leaf server periodically send partially-aggregated results to parent

% Caching strategy: 1. what content, 2. how long, 3. eviction

% Aggregation algorithm support: FedAvg \cite{mcmahan2017communication}, etc. As long as it satisfies associativity and commutativity.
% Optimization algorithm will be implemented by individual job frameworks.

% {\topic} communication protocol is 

% \subsection{Integration with Existing CDNs}
% 1. Relies on existing server selection algorithms for client request routing and load balancing for cost saving and performance.

% %TODO Reduction
\section{Implementation}
\label{sec:implementation}

{\name} is implemented in Python based on a microservice architecture. We use gRPC for communication between services, and Redis database for state management.
%TODO present overall workflow, some descriptions are in the code repo

%TODO scalability

% Given that concurrently operating schedulers
% may make conflicting scheduling decisions,
% we first divide the task to be scheduled among scheduling instances.
\section{Evaluation}
\label{sec:evaluation}

We evaluate the effectiveness of {\name} control plane on various metrics using simulated client traces from \cite{fedscale-icml22} and multi-tier jobs, with variance in demands and resource requirements. We choose two baseline implementations of collaborative ML resource selection: static partitioning, and pure random selection. We compare our system with these two baselines in terms of throughput, resource utilization, and job completion time. 

In addition, We study the performance of different scheduling algorithms in terms of throughput, resource utilization, job completion time, request completion time, execution completion time, and failure rate.

\subsection{Methodology}

\textbf{Static partitioning}: Every client is assigned to a fixed parameter server based on its attributes and server constraints. The number of assigned clients is proportional to the server's demand. There is no resource sharing among jobs since the partitioning is static.

\textbf{Pure random}: Every clients select a random parameter server to bind to. Parameter server conducts client selection based on its resource constraints, including both public and private constraints. Clients do not know which job is actively selecting clients prior to making binding request. This is the resource selection method implemented by Apple \cite{federated-personalization} and Meta \cite{MLSYS2022_a8bc4cb1}.

\textbf{Metrics}: Throughput, resource utilization, job completion time

\subsection{Results}

%TODO description: Ideal clients (always available), 5 jobs with different tiers

\begin{figure}[H]
    \centering
    \begin{subfigure}[b]{0.3\textwidth}
        \includegraphics[width=1\textwidth]{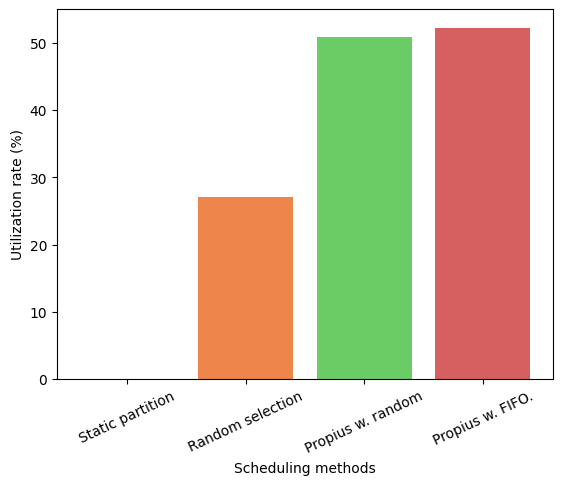}
        \caption{Client number: 1200}
        \label{fig:resource_util_1200}
    \end{subfigure}
    \begin{subfigure}[b]{0.3\textwidth}
        \includegraphics[width=1\textwidth]{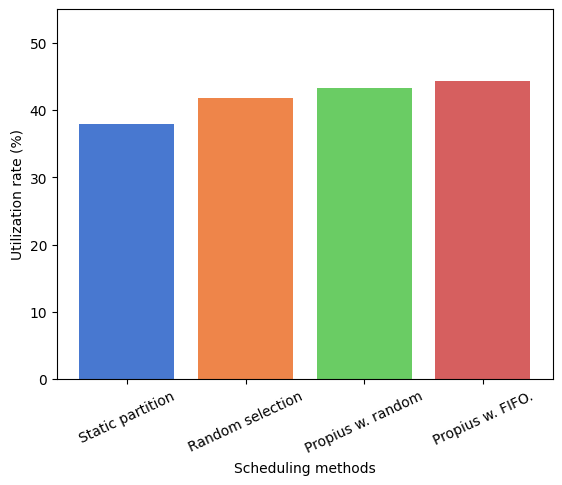}
        \caption{Client number: 1500}
        \label{fig:resource_util_1500}
    \end{subfigure}
    \begin{subfigure}[b]{0.3\textwidth}
        \includegraphics[width=1\textwidth]{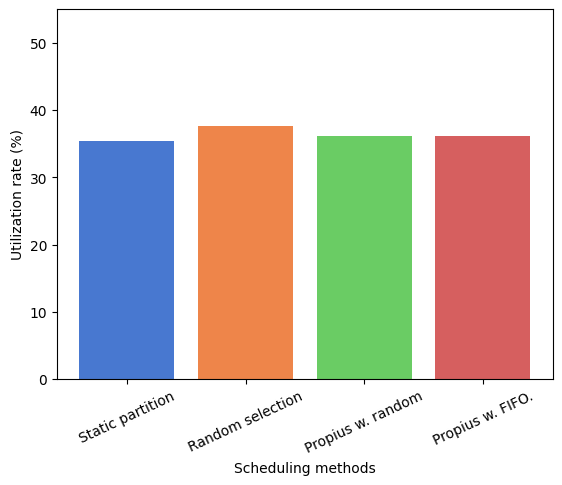}
        \caption{Client number: 1800}
        \label{fig:resource_util_1800}
    \end{subfigure}
    \caption{Resource utilization of different scheduling method. Resource utilization is 
    defined as the average client partitipation time (contributing to a job) divided by the overall 
    job completion time.}
    \label{fig:resource_util}
\end{figure}

\textbf{Resource utilization}: Through resource sharing, random selection and {\name} selection outperforms static partition in terms of resource utilization, especially when available resources are scarce (Fig. \ref{fig:resource_util}). 
When there are 1500 active clients, random selection achieves a $1.1\times$ improvement in resource utilization compared to static partition, while {\name} selection with random policy achieves a $1.14\times$ improvement. {\name}'s additional contribution to resource utilization can be attributed to its centrally managed pairing process of jobs with resource constraints and clients, as well as dynamically adjusting client allocation based on job demand.
The resource-sharing benefit is more pronounced when the available clients are scarce. With 1200 active clients, static partition selection cannot meet the demand of individual jobs, resulting in failed runs. {\name}'s efficient management of resources allows it to improve client resource utilization by $1.88\times$ compared to random selection under pure random scheme.
As the number of clients increases, every selection method achieves relatively similar resource utilization rate.

\begin{figure}[H]
    \centering
    \begin{subfigure}[b]{0.3\textwidth}
        \includegraphics[width=1\textwidth]{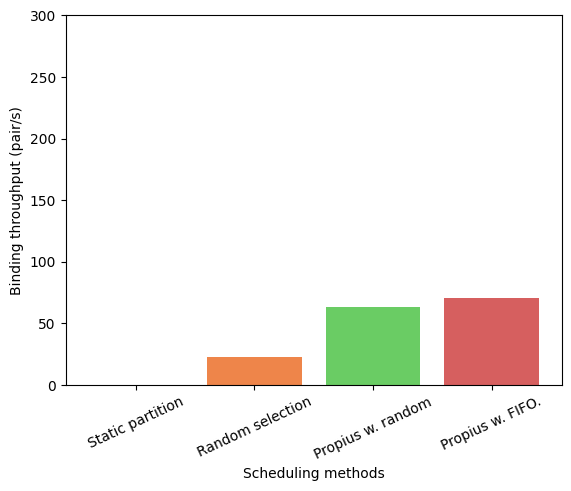}
        \caption{Client number: 1200}
        \label{fig:throughput_1200}
    \end{subfigure}
    \begin{subfigure}[b]{0.3\textwidth}
        \includegraphics[width=1\textwidth]{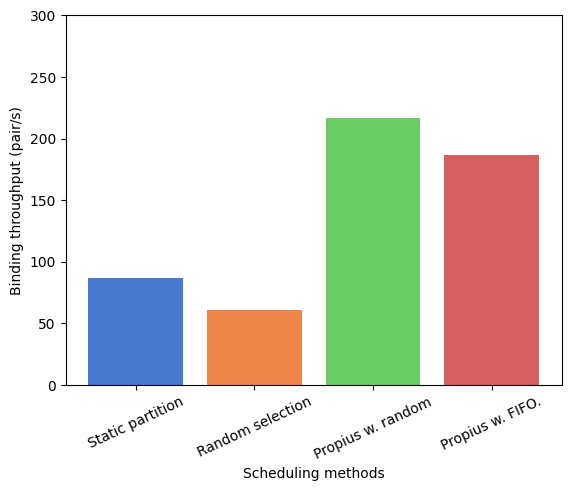}
        \caption{Client number: 1500}
        \label{fig:throughput_1500}
    \end{subfigure}
    \begin{subfigure}[b]{0.3\textwidth}
        \includegraphics[width=1\textwidth]{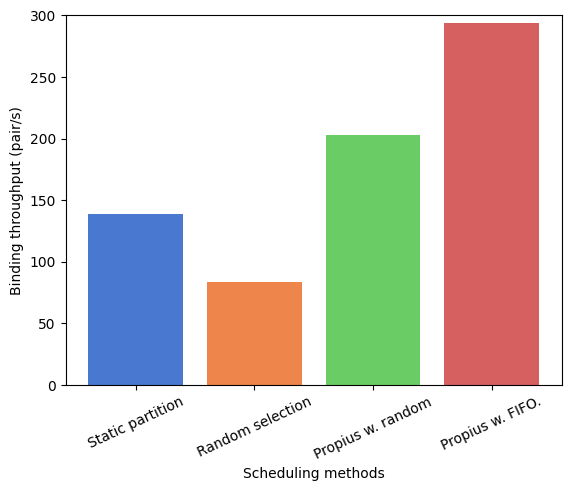}
        \caption{Client number: 1800}
        \label{fig:throughput_1800}
    \end{subfigure}
    \caption{Binding throughput of different scheduling methods. Throughput is defined as the number 
    of client-job binding per second.}
    \label{fig:throughput}
\end{figure}

\textbf{Throughput}: When there are ample resources available for each job, static partition method outperforms random selection in terms of binding throughput as each client is dedicated to a specific workload, removing the pairing handshake process in random selection. Yet {\name} performs the best, as is shown in Fig. \ref{fig:throughput}. Under random policy, {\name} achieves a $2.76\times$ improvement in throughput compared to random selection at best. This is because {\name}'s central scheduling scheme could efficiently maps available clients to active jobs in time, while clients have to execute multiple handshakes to find suitable jobs in random selection.
Also, Fig. \ref {fig:throughput} also shows the robustness of {\name} under varying loads.

\textbf{Job completion time (JCT)}: With the benefit of resource sharing and efficient pairing, {\name} reduces the average JCT by at most $1.264\times$. 
It is also interesting to notice that random selection outperforms static partition method in JCT, even though it has a lower binding throughput. This can be explained by the fact that resource sharing prevents client resources with better performance being monopolized by certain jobs, and poor-performing clients being stragglers for the entire duration of a job workload.

\begin{figure}[H]
    \centering
    \begin{subfigure}[b]{0.3\textwidth}
        \includegraphics[width=1\textwidth]{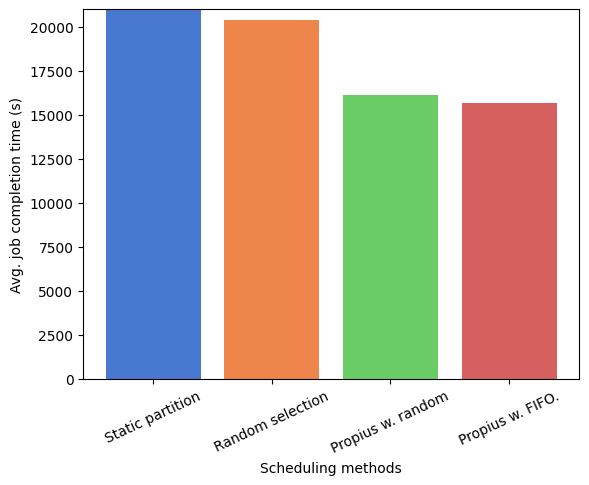}
        \caption{Client number: 1200}
        \label{fig:jct_1200}
    \end{subfigure}
    \begin{subfigure}[b]{0.3\textwidth}
        \includegraphics[width=1\textwidth]{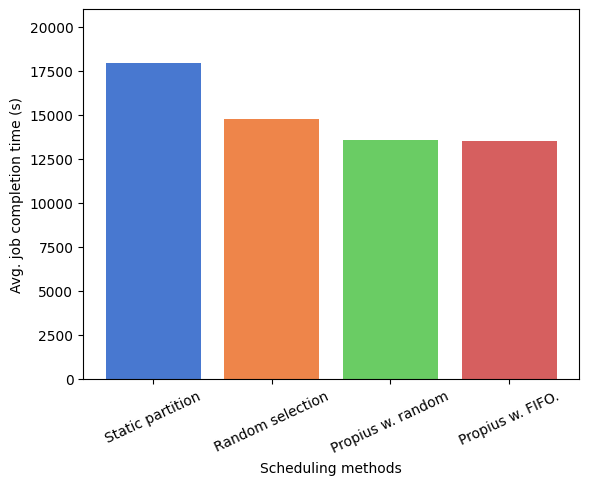}
        \caption{Client number: 1500}
        \label{fig:jct_1500}
    \end{subfigure}
    \begin{subfigure}[b]{0.3\textwidth}
        \includegraphics[width=1\textwidth]{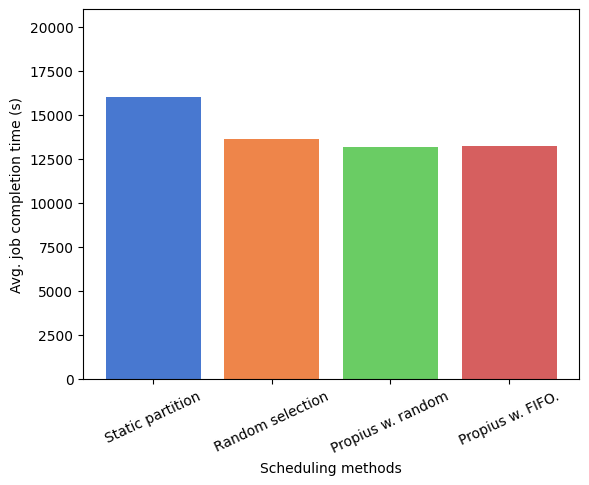}
        \caption{Client number: 1800}
        \label{fig:jct_1800}
    \end{subfigure}
    \caption{Average job completion time of different scheduling methods}
    \label{fig:jct}
\end{figure}

% Request completion time:
% 1500 ideal clients
% \begin{figure}[H]
%     \centering
%     \includegraphics[width=0.5\textwidth]{figures/eval_control_rct.png}
%     \caption{Average job request completion time of different scheduling methods}
%     \label{fig:rct}
% \end{figure}

% Round execution completion time:
% 1500 ideal clients
% \begin{figure}[H]
%     \centering
%     \includegraphics[width=0.5\textwidth]{figures/eval_control_ect.png}
%     \caption{Average job round execution completion time of different scheduling methods}
%     \label{fig:ect}
% \end{figure}

% Failure rate
%TODO

%TODO add results for 1200

% \subsubsection{Maintaining full client states vs. partial client states}

% \subsubsection{Study of different algorithms}
% %TODO

% \subsection{Data Plane}

% use random scheduler

% Metrics: Throughput of model down/up, JCT

% Taking into account network latency and throughput, running real workloads

% 1. Effectiveness of CDN in job convergence time

% \subsection{Scalability} % Combined

% Metrics: task ramp up latency, JCT

\section{Conclusion}
\label{sec:conclusion}

In this paper, we propose a novel {\topic} system, {\name}, that can adapt to the heterogeneity of client machines, and efficiently manage and control the computation flow between ML jobs and edge resources in a scalable fashion.
{\name} is comprised of a control plane and a data plane. The control plane enables efficient resource sharing among multiple {\topic} jobs and supports various resource sharing policies, while the data plane improves the scalability of {\topic} model sharing and result collection. Evaluations show that {\name} outperforms existing resource management techniques and frameworks in terms of resource utilization (up to $1.88\times$), throughput (up to $2.76\times$), and job completion time (up to $1.26\times$).
    %TODO data
% \input{pages/8-related.tex}
\label{EndOfPaper}

% References
\bibliographystyle{plain}
\bibliography{references/references}

\clearpage
\appendix

\end{document}